\documentclass[11pt,twoside]{article}

\usepackage{asp2006}

\usepackage{epsf}

\usepackage{psfig}

\usepackage{lscape}

\begin{document}

\title{The PSU/TCfA Search for Planets Around Evolved Stars:\\ Bisector Analysis of Activity of a Sample of Red Giants.}

\begin{quote}

Grzegorz Nowak$^1$ and Andrzej Niedzielski$^{1,2}$

{\itshape $^1$Torun Centre for Astronomy, Nicolaus Copernicus University,

ul. Gagarina 11, 87-100 Torun, Poland}

{\itshape $^2$Department of Astronomy and Astrophysics, Pennsylvania State

University, 525 Davey Laboratory, University Park, PA 16802}

\end{quote}

\begin{abstract}
Searches for planets around evolved G-K subgiant and giant stars are essential for developing general understanding of planet formation and evolution of the planetary systems. Precise radial velocity (RV) measurements of giants have lead to the discovery of ten planets around such star. However, the long period radial velocity variations of red giants may also have other than planetary nature. Non-radial oscillations or rotational modulation due to starspots can also induce RV variations, thereby mimicking the gravitational influence of low-mass companions. In this work we present bisector analysis of five carefully selected lines for two stars from our survey.

\end{abstract}

\section*{Introduction and observations}

Changes in line shapes arising from stellar atmospheric motion or from light contamination from unseen stellar companion can mimic small radial velocity variations at the spectral resolution 50,000 - 70,000 typically utilized for planet searches. Therefore it is important (especially in case of giant stars) to investigate whether the observed radial velocity curve are caused by a shift of the spectral lines as a whole or by a change in the symmetry of the spectral lines.

The observational material and reduction are described in Niedzielski \& Wolszczan (this volume).

\section*{Bisector analysis and results}

A basic tool to study the origin of RV variations derived on stellar spectra is the analysis of the shapes of spectral lines via line bisectors \citet{Gray1983}.

We used for bisector analysis lines with wavelengths longer than 6600 \AA where the Iodine cell has negligible spectrum. We computed line bisectors for 5 strong, unblended spectral features of a moderate intensity, which were located close to the center of Echelle orders: Cr I 6630.03, Ni I 6646.38, Ca I 6717.70, Fe I 6750.20, and Ni I 6767.84. All these lines show well defined bisectors.

The changes in the spectral line bisector were quantified using the bisector velocity span ($BVS$), which is simply the velocity difference between upper and lower point of the line bisector ($BVS = v_3 - v_2$) and bisector curvature ($BC$) which is the difference of the velocity span of the upper half of the bisector and the lower half ($BC = (v_3 - v_2) - (v_2 - v_1)$). It is important to examine both $BVS$ and $BC$ because it is possible for a star to show variations in one of these quantities only. In choosing the span points, it is important to avoid the wings and cores of the spectral line where the error of the bisector measurements are large. For our span measurements we chose $v_1=0.29$, $v_2=0.57$, and $v_3=0.79$ in the term of the line depth at the line core. Using the bisector measurements of all 5 spectral lines we compute an average velocity span and curvature after subtracting the mean value for each spectral line.

\begin{figure}[!h]

\begin{center}

\plotone{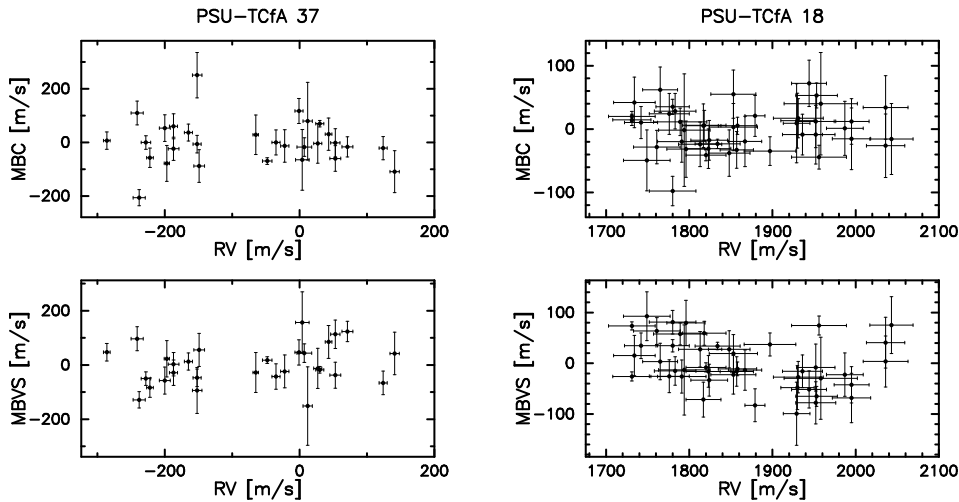}

\end{center}

\caption{Mean BVS and mean BC vs. RV for PSU-TCfA 37 (left) and PSU-TCfA 18 (right).}\label{37and18}

\end{figure}

In Figure \ref{37and18} we present mean bisector velocity span (MBVS), mean bisector curvature (MBC) for PSU-TCfA 37 (left panel) and PSU-TCfA 18 (right panel) as a function radial velocity (RV). Uncertainties in the derived values of the MBVS and MBC were estimated as standard deviations of the mean. The correlation coefficients were find to be r = 0.22 $\pm$ 0.05 for MBVS and r = -0.44 $\pm$ 0.08 for MBC for PSU-TCfA 37 and r = 0.32 $\pm$ 0.10 for MBVS and r = 0.07 $\pm$ 0.02 for PSU-TCfA 18. It is clear that they are not correlated with radial velocities what supports the planetary mass companion hypothesis.

\acknowledgements We thank the HET resident astronomers and telescope operators for cooperation. We acknowledge the financial support from the MNiSW through grant 1P03D 007 30. GN is a recipient of a graduate stipend of the Chairman of the Polish Academy of Sciences.


\end{document}